\documentclass[conference]{IEEEtran}

\usepackage{amsmath,amsfonts,amssymb,amsthm}
\usepackage{mathrsfs}
\usepackage{cite}
\usepackage{graphicx}
\usepackage{booktabs}
\usepackage{algorithmic}
\usepackage{algorithm}
\usepackage{balance}
\usepackage{placeins}

\usepackage{graphicx}
\usepackage{subfig}
\usepackage[font=footnotesize,labelfont=bf,labelsep=colon]{caption}

\usepackage{multirow} 
\usepackage{authblk}

\theoremstyle{definition}

\begin{document}

\title{Spectral-Convolutional Factorized Transformer for Joint Multipath Parameter Estimation: Applications to mmWave Positioning and Radio SLAM}

\author[]{Yaya Etiabi}
\author[]{Wessam Ajib}
\author[]{Halima Elbiaze}
\affil[]{Department of Computer Science,
Université du Québec à Montréal (UQAM), Montréal, Canada
\authorcr Emails: {\{etiabi.yaya, elbiaze.halima, ajib.wessam\}@uqam.ca}}


\maketitle

\begin{abstract}
Future 6G wireless networks are expected to evolve toward AI-native systems that support intelligent, environment-aware services, including positioning and mapping. Current research has already integrated AI into several aspects of communication, particularly wireless channel estimation. However, most learning-based channel estimators focus on communication-oriented objectives rather than explicitly estimating multipath parameters such as delay, angle of arrival (AoA), and angle of departure (AoD). These parameters provide essential geometric information for downstream tasks including positioning and radio simultaneous localization and mapping (SLAM). In this work, we propose a Spectral-Convolutional Factorized Transformer (SCFT) for joint multipath component (MPC) estimation from noisy complex channel estimates and physical beam-pair descriptors. SCFT combines local spectral convolution, permutation-equivariant beam-set attention, and a query-based Transformer decoder to detect and classify line-of-sight and non-line-of-sight paths and estimate their observed delays, AoAs, and AoDs. Experiments on realistic ray-tracing data show that SCFT reduces the generalized optimal sub-pattern assignment error by 13.1\% compared with ResNet, the strongest competing baseline. We further demonstrate this advantage in independent snapshot positioning and radio SLAM, where SCFT reduces position RMSE by 18.5\% and 65.0\%, respectively, relative to ResNet. These findings highlight the importance of explicitly recovering multipath geometry, as improvements in MPC estimation can translate into substantial gains in downstream tasks.
\end{abstract}

\begin{IEEEkeywords}
Channel estimation, multipath parameter estimation, millimeter-wave, MIMO-OFDM,
factorized transformer, positioning, radio SLAM.
\end{IEEEkeywords}

\section{Introduction}


The proliferation of millimeter-wave (mmWave) multiple-input
multiple-output (MIMO) orthogonal frequency-division multiplexing
(OFDM) systems in 5G-Advanced and future 6G networks creates
opportunities beyond high-rate communication
\cite{dai2026tutorial}. The large bandwidth and massive
antenna arrays of mmWave MIMO systems enable high-accuracy
radio-based positioning, tracking, and environmental mapping.
These capabilities are central to the 6G vision of integrated sensing, localization, and communication~\cite{guo2025integrated}, where shared radio resources support data transmission and environmental perception.
In parallel, the AI-native network paradigm envisioned in 6G
\cite{hoydis2021toward} motivates the use of deep learning to
overcome the limitations of classical model-based algorithms
in the face of dynamic, multipath-rich propagation. By
exploiting the geometric relationship between propagation
paths and the physical environment, joint estimation of
multipath parameters, such as delays, angles of
arrival/departure (AoA/AoD), and Doppler shifts, supports
downstream tasks such as precise positioning and radio
simultaneous localization and mapping (SLAM)
\cite{rastorgueva2024millimeter}.
A central challenge for these applications is to reliably
estimate multipath delays and angles from noisy,
high-dimensional beamformed channel observations across
subcarriers and beam pairs.
Classical model-based methods such as MUSIC~\cite{schmidt1986multiple}, SAGE~\cite{SAGE}, and ESPRIT~\cite{ESPRIT} provide established approaches to multipath parameter estimation. Their application to joint delay and angular recovery can involve multidimensional searches or additional estimation stages, with performance potentially degrading under challenging propagation and noise conditions. Compressive sensing (CS) approaches exploit angular-delay sparsity \cite{OMP,tian2026gsmomp}, yet grid-based methods suffer from basis mismatch, while grid-free approaches can incur substantial computational overhead. Tensor decomposition, e.g., via canonical polyadic models~\cite{gong2025deep}, offers a principled multidimensional framework, but conventional alternating least-squares solvers can be sensitive to initialization and require many iterations to converge.
 
In parallel, deep learning has been extensively investigated for wireless channel processing, including channel reconstruction~\cite{soltani2019deep}, equalization~\cite{jin2026equalization}, and channel state information (CSI) feedback~\cite{wang2019deep}. However, these communication-oriented approaches generally do not explicitly recover the geometric multipath parameters, such as delay, AoA, and AoD, needed for positioning and radio SLAM. More recent studies have investigated learning-based methods for multipath parameter estimation~\cite{zhou2026multipath}, including Transformer-based delay estimation~\cite{ott2024multipath}. Transformers~\cite{vaswani2017attention}, despite their success in other domains, remain relatively underexplored for joint multipath geometric parameter estimation.
A direct application of self-attention would first flatten the antenna, subcarrier, and time dimensions into a single sequence. This leads to quadratic complexity in the resulting sequence length and becomes costly for large MIMO-OFDM observations. 
Such a formulation also does not explicitly exploit the
separable spatial and frequency structure of multipath
propagation.

To bridge this gap, we propose the Spectral-Convolutional
Factorized Transformer (SCFT) for joint channel-parameter
estimation directly from complex beamformed CSI. Rather than flattening frequency and beam-pair observations into a generic sequence, SCFT uses local convolution along the ordered spectral axis and set attention across the unordered beam-pair axis before decoding path-level representations. This design follows the separable frequency and directional structure induced by multipath propagation. The resulting estimator directly predicts delay and AoA/AoD for positioning and Radio SLAM.
Accordingly, the contributions of this paper are summarized as follows:

\begin{itemize}
    \item We propose SCFT for joint path-level channel-parameter estimation in mmWave MIMO-OFDM systems, combining spectral convolution, beam-set attention, learnable multipath component (MPC) queries, and a permutation-invariant assignment loss.
    \item We evaluate SCFT against representative classical and learning-based estimators and conduct ablation studies to assess the contributions of its architectural components.
    \item We evaluate the estimated MPCs in independent snapshot positioning and full causal radio SLAM over repeated noisy trajectory realizations.
\end{itemize}
The remainder of the paper is organized as follows.
Section~II presents the system model and problem formulation.
Section~III describes the proposed SCFT architecture.
Section~IV presents the performance evaluation and Section~V concludes the paper.

\section{System Model and Problem Formulation}
\label{sec:sysmodel}
\subsection{MIMO-OFDM Channel Model}

We consider a downlink mmWave MIMO-OFDM system comprising a base station (BS) and a user equipment (UE) equipped with uniform planar arrays of $N_{\mathrm a}^{\mathrm{TX}}$ and $N_{\mathrm a}^{\mathrm{RX}}$ elements, respectively. The channel is assumed static within each snapshot $t$ and is observed over $N_{\mathrm{sc}}$ subcarriers with spacing $\Delta f$. Its frequency response at subcarrier $k\in\{0,\ldots,N_{\mathrm{sc}}-1\}$ is modeled as a superposition of $L_t$ multipath components (MPCs)~\cite{channel_model}:
\begin{equation}
\mathbf H_t[k]=\sum_{\ell=1}^{L_t}\alpha_{t,\ell}
e^{-\mathrm{i}2\pi k\Delta f\widetilde\tau_{t,\ell}}
\mathbf a_{\mathrm r}(\phi_{t,\ell},\epsilon^{\mathrm r}_{t,\ell})
\mathbf a_{\mathrm t}^{H}(\theta_{t,\ell},\epsilon^{\mathrm t}_{t,\ell}),
\label{eq:channel_model}
\end{equation}
where $\alpha_{t,\ell}$ is the complex path gain, $\phi_{t,\ell}$ and $\theta_{t,\ell}$ are the arrival and departure azimuths, and $\epsilon^{\mathrm r}_{t,\ell}$ and $\epsilon^{\mathrm t}_{t,\ell}$ are the corresponding elevation angles. The functions $\mathbf a_{\mathrm r}$ and $\mathbf a_{\mathrm t}$ denote the receive and transmit array responses. The observed delay $\widetilde\tau_{t,\ell}=\tau_{t,\ell}+b_t$ includes the propagation delay $\tau_{t,\ell}$ and receiver clock bias $b_t$.
Let $\mathcal S_t$ denote the set of selected beam pairs, with $|\mathcal S_t|\leq S$. For each pair $p\in\mathcal S_t$, the receive and transmit beamforming vectors are denoted by $\mathbf w_{\iota_{t,p}}$ and $\mathbf f_{\kappa_{t,p}}$, respectively. The noisy beamformed channel observation is given by
\begin{align}
\widehat h_{t,p}[k]
&=\mathbf w_{\iota_{t,p}}^{H}\mathbf H_t[k]\mathbf f_{\kappa_{t,p}}
+\widetilde n_{t,p}[k] \nonumber\\
&=\sum_{\ell=1}^{L_t}\alpha_{t,\ell}g_{t,p,\ell}
e^{-\mathrm{i}2\pi k\Delta f\widetilde\tau_{t,\ell}}
+\widetilde n_{t,p}[k],
\label{eq:beam_csi}
\end{align}
where $g_{t,p,\ell}=(\mathbf w_{\iota_{t,p}}^{H}\mathbf a_{\mathrm r}(\phi_{t,\ell},\epsilon^{\mathrm r}_{t,\ell}))(\mathbf a_{\mathrm t}^{H}(\theta_{t,\ell},\epsilon^{\mathrm t}_{t,\ell})\mathbf f_{\kappa_{t,p}})$ is the beam-pair response to path $\ell$, and $\widetilde n_{t,p}[k]$ denotes the additive noise. The resulting observations exhibit a frequency-dependent phase progression within each beam pair and directional variations across the selected beam pairs.

\subsection{Complex CSI and Beam-Pair Observation}

We normalize the complex beamformed channel observations at each snapshot by their root-mean-square magnitude over the selected beam pairs:
\begin{equation}
 \rho_t=\left(\frac{1}{|\mathcal S_t|N_{\mathrm{sc}}}
 \sum_{p\in\mathcal S_t}\sum_{k=0}^{N_{\mathrm{sc}}-1}
 |\widehat h_{t,p}[k]|^2\right)^{1/2},
 \label{eq:csi_rms}
\end{equation}
The normalized observations are represented by their real and imaginary components:
\begin{equation*}
 \mathbf x_{t,p}[k]=\rho_t^{-1}\!\left[\Re\{\widehat h_{t,p}[k]\},
 \Im\{\widehat h_{t,p}[k]\}\right]^{T}.
\end{equation*}
This normalization removes a common amplitude scale while preserving relative amplitudes and phases.

Each beam pair is represented by a real-valued descriptor constructed from its receive and transmit codebook vectors:
\begin{equation}
 \mathbf d_{t,p}=\left[\Re\{\mathbf w_{\iota_{t,p}}\}^{T},
 \Im\{\mathbf w_{\iota_{t,p}}\}^{T},
 \Re\{\mathbf f_{\kappa_{t,p}}\}^{T},
 \Im\{\mathbf f_{\kappa_{t,p}}\}^{T}\right]^{T}.
 \label{eq:beam_descriptor}
\end{equation}
The estimator takes $(\mathbf X_t,\mathbf D_t,\mathbf m_t)$ as input, where $\mathbf X_t\in\mathbb R^{S\times N_{\mathrm{sc}}\times2}$ contains the normalized channel observations, $\mathbf D_t\in\mathbb R^{S\times2(N_{\mathrm a}^{\mathrm{RX}}+N_{\mathrm a}^{\mathrm{TX}})}$ contains the beam descriptors, and $\mathbf m_t\in\{0,1\}^{S}$ indicates valid beam pairs. Since beam pairs form an unordered set, the estimator must be invariant to their ordering. For any row-permutation matrix $\mathbf P\in\{0,1\}^{S\times S}$, the estimator $f_{\boldsymbol\Theta}$ therefore satisfies
\begin{equation}
 f_{\boldsymbol\Theta}(\mathbf P\mathbf X_t,
 \mathbf P\mathbf D_t,\mathbf P\mathbf m_t)
 =f_{\boldsymbol\Theta}(\mathbf X_t,\mathbf D_t,\mathbf m_t).
 \label{eq:beam_permutation}
\end{equation}

\subsection{Unordered MPC Set Estimation}

We formulate MPC estimation as the prediction of an unordered set of path parameters. At snapshot $t$, the target set is
\begin{equation}
 \mathcal Z_t=\left\{z_{t,\ell}=\left(c_{t,\ell},
 \widetilde\tau_{t,\ell},\phi^{\mathrm B}_{t,\ell},
 \theta^{\mathrm G}_{t,\ell}\right)\right\}_{\ell=1}^{L_t},
 \label{eq:true_set}
\end{equation}
where $c_{t,\ell}\in\{0,1\}$ denotes LoS and NLoS paths, respectively, while $\phi^{\mathrm B}_{t,\ell}$ and $\theta^{\mathrm G}_{t,\ell}$ are the receiver-frame AoA and global-frame AoD.
 We consider single-reflection NLoS paths and exclude path gains and Doppler from estimation.

Using $M\geq L_t$ candidate queries, the estimator predicts
\begin{equation}
 \widehat{\mathcal Z}_t=\left\{\widehat z_{t,i}=\left(\widehat p_{t,i},
 \widehat{\boldsymbol\pi}_{t,i},\widehat{\widetilde\tau}_{t,i},
 \widehat\phi^{\mathrm B}_{t,i},\widehat\theta^{\mathrm G}_{t,i}
 \right)\right\}_{i=1}^{M},
 \label{eq:pred_set}
\end{equation}
where $\widehat p_{t,i}$ is the path-existence probability and $\widehat{\boldsymbol\pi}_{t,i}$ is the two-class distribution. Since MPCs are unordered, predicted queries are matched to target paths through
\begin{equation}
 \xi_t^{\star}=\arg\min_{\xi\in\mathfrak I_{L_t,M}}
 \sum_{\ell=1}^{L_t}C\!\left(z_{t,\ell},
 \widehat z_{t,\xi(\ell)}\right),
 \label{eq:hungarian_assignment}
\end{equation}
where $\mathfrak{I}_{L_t,M}$ is the set of injective mappings from target paths to queries. The assignment cost and permutation-invariant training loss are defined in Section~III-D.


\section{Proposed Spectral-Convolutional Factorized Transformer}
\label{sec:proposed}

\begin{figure}
    \centering
    \includegraphics[width=\linewidth]{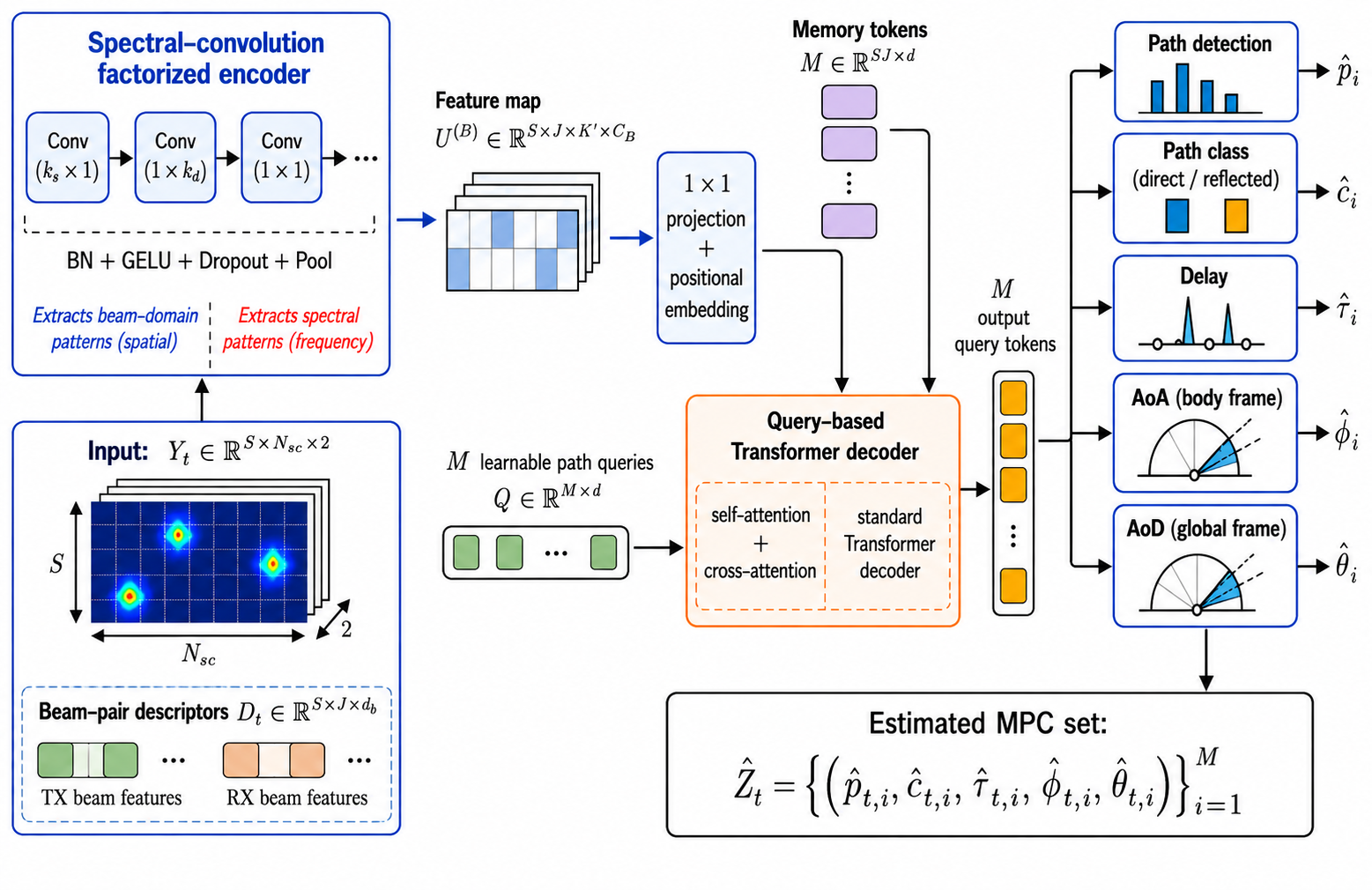}
    \caption{Overview of the proposed Spectral-Convolutional Factorized Transformer (SCFT) framework for joint multipath component estimation.}
    \label{fig:architecture}
\end{figure}

As illustrated in Fig.~\ref{fig:architecture}, SCFT processes the ordered frequency response within each beam pair using local spectral convolution and combines information across unordered beam pairs through set attention. Let $\mathbf Z_t^{(0)}=\mathcal T_{\mathrm{CSI}}(\mathbf X_t,\mathbf D_t,\mathbf m_t)$ denote the tokenized input. Each of the $N_b$ encoder blocks computes
\begin{equation}
 \mathbf Z_t^{(n)}=\mathcal A_{\mathrm{pair}}^{(n)}
 \!\left(\mathcal C_{\mathrm{spec}}^{(n)}
 \!\left(\mathbf Z_t^{(n-1)}\right),\mathbf m_t\right),
 \quad n=1,\ldots,N_b,
 \label{eq:factorized_mapping}
\end{equation}
where $\mathcal T_{\mathrm{CSI}}$ denotes complex-CSI tokenization, $\mathcal C_{\mathrm{spec}}$ spectral convolution, and $\mathcal A_{\mathrm{pair}}$ beam-pair set attention.

\subsection{Complex Spectral Tokens}

Each beam-pair CSI sequence is divided into $J$ contiguous spectral patches. Let $\mathbf x_{t,p}^{(j)}$ contain the real and imaginary samples of patch $j\in\{1,\ldots,J\}$. A shared patch encoder produces a $d$-dimensional token, augmented with beam-pair and spectral-position embeddings:
\begin{equation}
 \mathbf z^{(0)}_{t,p,j}=\mathcal T_{\mathrm{patch}}
 (\mathbf x_{t,p}^{(j)})+\mathbf W_d\mathbf d_{t,p}+\boldsymbol\eta_j,
 \label{eq:spectral_token}
\end{equation}
where $\mathbf W_d$ projects the beam-pair descriptor and $\boldsymbol\eta_j$ encodes the spectral position.
The complex exponential associated with each path in~\eqref{eq:beam_csi} exhibits a consistent phase progression across subcarriers. For an integer subcarrier lag $\nu$,
\begin{equation}
 \frac{e^{-\mathrm{i}2\pi(k+\nu)\Delta f\widetilde\tau}}
 {e^{-\mathrm{i}2\pi k\Delta f\widetilde\tau}}
 =e^{-\mathrm{i}2\pi\nu\Delta f\widetilde\tau},
 \label{eq:local_phase_progression}
\end{equation}
This structure motivates shared convolution along the spectral dimension. Within each beam pair $p$, encoder block $n$ applies
\begin{align}
 \mathcal G_{\mathrm s}(\mathbf Y)&=\mathcal A_c\!\left(
 \mathrm{Conv}_{1}(\mathrm{GELU}(\mathrm{Conv}_{k_s}(\mathrm{LN}(\mathbf Y))))
 \right),\nonumber\\
 \mathbf V^{(n)}_{t,p}&=\mathbf Z^{(n-1)}_{t,p,:,:}
 +\mathcal G_{\mathrm s}(\mathbf Z^{(n-1)}_{t,p,:,:}),\nonumber\\
 \widetilde{\mathbf Z}^{(n)}_{t,p,:,:}&=\mathbf V^{(n)}_{t,p}
 +\mathrm{FFN}_{\mathrm s}(\mathrm{LN}(\mathbf V^{(n)}_{t,p})),
 \label{eq:spectral_mixer}
\end{align}
where $\mathrm{Conv}_{k_s}$ is a one-dimensional convolution of width $k_s$, $\mathrm{Conv}_{1}$ is a pointwise projection, $\mathcal A_c$ is a channel-wise affine transformation, and $\mathrm{FFN}_{\mathrm s}$ is the spectral feed-forward network.

\subsection{Permutation-Equivariant Beam-Pair Attention}

At each spectral patch $j$, the beam-pair tokens $\widetilde{\mathbf Z}^{(n)}_{t,:,j,:}$ interact through masked self-attention followed by feed-forward processing:
\begin{align}
 \mathbf U^{(n)}_{t,j}&=\mathrm{LN}
 (\widetilde{\mathbf Z}^{(n)}_{t,:,j,:}),\nonumber\\
 \overline{\mathbf Z}^{(n)}_{t,:,j,:}&=\widetilde{\mathbf Z}^{(n)}_{t,:,j,:}
 +\mathrm{MHA}(\mathbf U^{(n)}_{t,j},\mathbf U^{(n)}_{t,j},
 \mathbf U^{(n)}_{t,j};\mathbf m_t),\nonumber\\
 \mathbf Z^{(n)}_{t,:,j,:}&=\overline{\mathbf Z}^{(n)}_{t,:,j,:}
 +\mathrm{FFN}_{\mathrm p}(\mathrm{LN}(\overline{\mathbf Z}^{(n)}_{t,:,j,:})),
 \label{eq:pair_attention}
\end{align}
where $\mathrm{FFN}_{\mathrm p}$ is the beam-pair feed-forward network and $\mathbf m_t$ masks padded beam pairs. Since no positional encoding is assigned to beam-pair rows, the attention block is permutation-equivariant:
$\mathcal A_{\mathrm{pair}}(\mathbf P\mathbf Z,\mathbf P\mathbf m_t)
=\mathbf P\mathcal A_{\mathrm{pair}}(\mathbf Z,\mathbf m_t)$.
Spectral convolution applies the same operations independently to each beam pair, while decoder cross-attention is invariant to a common permutation of its keys and values. Consequently, SCFT satisfies~\eqref{eq:beam_permutation}.

\subsection{Full-Memory MPC Query Decoder}

The output of the $N_b$ encoder blocks is layer-normalized and reshaped into a memory $\mathbf M_t\in\mathbb R^{SJ\times d}$, preserving all spectral and beam-pair tokens. The decoder uses $M$ learnable MPC queries $\mathbf Q^{(0)}\in\mathbb R^{M\times d}$. At each of its $R$ layers, self-attention among queries is followed by cross-attention with the encoder memory and feed-forward processing:
\begin{align}
 \mathbf U_Q^{(r)}&=\mathrm{LN}(\mathbf Q^{(r-1)}),\nonumber\\
 \overline{\mathbf Q}^{(r)}&=\mathbf Q^{(r-1)}+\mathrm{MHA}
 (\mathbf U_Q^{(r)},\mathbf U_Q^{(r)},\mathbf U_Q^{(r)}),\nonumber\\
 \widetilde{\mathbf Q}^{(r)}&=\overline{\mathbf Q}^{(r)}+
 \mathrm{MHA}(\mathrm{LN}(\overline{\mathbf Q}^{(r)}),
 \mathbf M_t,\mathbf M_t;\overline{\mathbf m}_t),\nonumber\\
 \mathbf Q^{(r)}&=\widetilde{\mathbf Q}^{(r)}+
 \mathrm{FFN}(\mathrm{LN}(\widetilde{\mathbf Q}^{(r)})),
 \label{eq:query_decoder}
\end{align}
where $\overline{\mathbf m}_t\in\{0,1\}^{SJ}$ extends each beam-pair mask entry across its $J$ spectral tokens. The decoded queries produce MPC predictions in parallel, following the set-prediction approach of~\cite{carion2020end}.
Each query is mapped to existence, path-class, delay, AoA, and AoD predictions through separate output heads. For $\psi\in\{\phi^{\mathrm B},\theta^{\mathrm G}\}$, angles are represented by unit vectors $\widehat{\mathbf u}_{\psi,t,i}=[\cos\widehat\psi_{t,i},\sin\widehat\psi_{t,i}]^T$ to avoid angular wrap-around discontinuities. The observed delay is standardized as $\bar{\tau}=(\widetilde{\tau}-\mu_{\tau})/\sigma_{\tau}$, where $\mu_{\tau}$ and $\sigma_{\tau}>0$ are normalization constants. The predicted delay is recovered as $\widehat{\widetilde{\tau}}=\sigma_{\tau}\widehat{\bar{\tau}}+\mu_{\tau}$.

\subsection{Permutation-Invariant Set Objective}

The assignment in~\eqref{eq:hungarian_assignment} uses a cost combining delay, path-class, and angular discrepancies. With $\mathbf u_{\psi,t,\ell}=[\cos\psi_{t,\ell},\sin\psi_{t,\ell}]^T$ and standardized delay $\bar\tau$, this cost is defined as
\begin{align}
 C(z_{t,\ell},\widehat z_{t,i})={}&
 \lambda_{\tau}^{a}|\widehat{\bar\tau}_{t,i}-\bar\tau_{t,\ell}|
 -\lambda_c^{a}\log\widehat\pi_{t,i,c_{t,\ell}} \nonumber\\
 &+\lambda_{\phi}^{a}(1-\widehat{\mathbf u}_{\phi,t,i}^{T}
 \mathbf u_{\phi,t,\ell})
 +\lambda_{\theta}^{a}(1-\widehat{\mathbf u}_{\theta,t,i}^{T}
 \mathbf u_{\theta,t,\ell}),
 \label{eq:matching_cost}
\end{align}
where $\lambda_\tau^a$, $\lambda_c^a$, $\lambda_\phi^a$, and $\lambda_\theta^a$ are nonnegative assignment weights. The optimal assignment $\xi_t^\star$ defines the existence target $e_{t,i}$, set to 1 for matched queries and 0 otherwise.

For a batch $\mathcal B$ of snapshots, let $N_{\mathcal B}=\sum_{t\in\mathcal B}L_t$ and $i_{t,\ell}=\xi_t^\star(\ell)$. A superscript $\star$ on a prediction denotes evaluation at the matched query $i_{t,\ell}$. The training objective is
\begin{align}
 \mathcal L_{\mathcal B}={}&\frac{\lambda_{\mathrm{ex}}}{|\mathcal B|M}
 \sum_{t\in\mathcal B}\sum_{i=1}^{M}
 \mathrm{BCE}(\widehat p_{t,i},e_{t,i})\nonumber\\
 &+\frac{1}{N_{\mathcal B}}\sum_{t\in\mathcal B}\sum_{\ell=1}^{L_t}\Big[
 \lambda_c\mathrm{CE}(\widehat{\boldsymbol\pi}^{\star}_{t,\ell},c_{t,\ell})
 +\lambda_\tau\mathrm{SL1}(\widehat{\bar\tau}^{\star}_{t,\ell},
 \bar\tau_{t,\ell})\nonumber\\
 &\hspace{7mm}+\lambda_\phi(1-\widehat{\mathbf u}_{\phi,t,\ell}^{\star T}
 \mathbf u_{\phi,t,\ell})
 +\lambda_\theta(1-\widehat{\mathbf u}_{\theta,t,\ell}^{\star T}
 \mathbf u_{\theta,t,\ell})\Big].
 \label{eq:set_loss}
\end{align}
Here $\mathrm{SL1}$ denotes the Smooth-L1 loss, and the $\lambda$ coefficients are nonnegative loss weights. The existence loss applies to all queries, whereas classification and geometric losses apply only to matched queries. At inference, queries with $\widehat p_{t,i}\geq\gamma_{\mathrm{ex}}$ are retained, where $\gamma_{\mathrm{ex}}$ is the existence threshold.

\subsection{Computational Complexity}

A flat Transformer over $S\times J$ tokens incurs $\mathcal O(N_bS^2J^2d)$ attention complexity across $N_b$ blocks. By separating local spectral mixing from global beam-pair interactions, SCFT reduces the encoder complexity to \mbox{$\mathcal O\!\left(N_b(SJk_sd^2+JS^2d)\right)$}, excluding common projection and feed-forward costs. Unlike dense attention, whose cost grows quadratically with $J$, the factorized encoder scales linearly with the number of spectral patches for fixed $S$, $d$, and $k_s$. The full $SJ$-token memory remains accessible to all MPC queries, with decoder attention complexity $\mathcal O(R(M^2d+MSJd))$. Thus, SCFT avoids dense pair-frequency self-attention without sacrificing global access to the encoded observations.

\section{Performance Evaluation}
\label{sec:results}

\subsection{Experimental Setup}

To evaluate the performance of the proposed framework, we use realistic ray-tracing data representing an outdoor urban environment, specifically the O1$_{28}$ scenario from DeepMIMO v4~\cite{alkhateeb2019deepmimo}. Table~\ref{tab:sim_params} summarizes the experimental configuration. Training, validation, calibration, and test sets use distinct receiver locations, while positioning and SLAM are evaluated on a separate trajectory. For each observation, the $S$ RX--TX DFT beam pairs with the highest noisy average energy are selected and shared across estimators. The learning-based baselines (flat Transformer, ResNet, vanilla CNN, and spectral U-Net) use the same training objective and schedule as SCFT, with comparable parameter counts. SAGE~\cite{SAGE} and joint OMP~\cite{OMP} perform joint delay--angle searches, whereas MUSIC~\cite{schmidt1986multiple} and ESPRIT~\cite{ESPRIT} estimate delays from frequency-domain subspaces. All classical methods use the Akaike information criterion (AIC) for path-count estimation and delay-conditioned beamspace angular refinement.

For MPC evaluation, we use an existence threshold $\gamma_{\mathrm{ex}}=0.5$, where applicable, and the normalized geometric distance
\begin{equation}
d=\left[
\left(\frac{\Delta\widetilde\tau}{\sigma_\tau}\right)^2
+\left(\frac{\delta_\phi}{s_a}\right)^2
+\left(\frac{\delta_\theta}{s_a}\right)^2
\right]^{1/2},
\end{equation}
where $\sigma_\tau=101.99$ ns is the training-delay standard deviation, $s_a=10^\circ$, and $\delta_\phi,\delta_\theta$ are the minimum wrapped angular errors under the exact front/back array-response equivalence in their respective frames. We compute the generalized optimal sub-pattern assignment (GOSPA) metric~\cite{GOSPA} with $(p,c,\alpha)=(2,5,2)$ and class-agnostic, one-to-one assignment cost $\min(d,5)^2$. Matches with $d<5$ determine micro-averaged detection F1 and matched-path delay RMSE and angular MAEs. GOSPA is reported as mean and sample standard deviation.

\begin{table}[t]
\centering
\vspace*{5pt}
\caption{Experimental configuration.}
\label{tab:sim_params}
\scriptsize
\setlength{\tabcolsep}{2.2pt}
\renewcommand{\arraystretch}{1.02}
\begin{tabular}{ll}
\toprule
\textbf{Item} & \textbf{Value} \\
\midrule
Carrier; subcarrier spacing & $28$ GHz; $120$ kHz \\
Subcarriers; selected pairs & $N_{\mathrm{sc}}=768$; $S=32$ \\
BS/RX arrays; codebooks & $8\!\times\!8$/$4\!\times\!4$; Kronecker DFT \\
Train/val./cal./test samples & $35000/5000/5000/5000$ \\
SNR; MPC queries & $0$--$30$ dB; $M=5$ \\
SCFT encoder/decoder & $N_b=4$, $R=2$, $d=128$, 8 heads \\
FFN; dropout & 256; 0.1 \\
Spectral patches; convolution & length/stride 16 ($J=48$); $k_s=3$ \\
Optimizer; batch; epochs & AdamW; 64; 100 \\
Learning rate; weight decay & $3\!\times\!10^{-4}$; $10^{-4}$ \\
Schedule; Smooth-L1 & 5-epoch warmup, cosine to $10^{-6}$; $\beta=0.1$ \\
Loss weights $(\lambda_{\rm ex},\lambda_c,\lambda_\tau,\lambda_\phi,\lambda_\theta)$
& $(2,0.5,2,2,1)$ \\
Assignment weights $(\lambda_\tau^a,\lambda_\phi^a,\lambda_\theta^a,\lambda_c^a)$
& $(1,1,1,0.5)$ \\
\bottomrule
\end{tabular}
\end{table}

\subsection{MPC Estimation}

\begin{table*}[!t]
\centering
\vspace*{5pt}
\caption{MPC estimation performance. Delay and angular errors are evaluated over matched paths; GFLOPs are reported per sample.}
\label{tab:estimator_results}
\scriptsize
\setlength{\tabcolsep}{4.0pt}
\renewcommand{\arraystretch}{0.92}
\begin{tabular}{lrrrrrrr}
\toprule
\textbf{Method} & \textbf{Params (M)} & \textbf{GFLOPs} &
\textbf{GOSPA $\mu\!\pm\!\sigma$} & \textbf{F1} &
\textbf{Delay RMSE (ns)} & \textbf{AoA MAE ($^\circ$)} &
\textbf{AoD MAE ($^\circ$)} \\
\midrule
\textbf{SCFT} & 1.490 & 3.621 & $\mathbf{2.001\!\pm\!2.199}$ &
\textbf{0.942} & \textbf{16.80} & 5.31 & \textbf{2.26} \\
Flat Transformer & 1.491 & 4.081 & $2.797\!\pm\!2.487$ & 0.915 & 20.67 & 7.69 & 3.06 \\
ResNet & 1.520 & 3.614 & $2.303\!\pm\!2.301$ & 0.932 & 17.60 & 5.93 & 2.94 \\
Vanilla CNN & 1.520 & 3.612 & $5.900\!\pm\!2.333$ & 0.777 & 37.35 & 18.82 & 4.64 \\
Spectral U-Net & 1.563 & 3.583 & $5.341\!\pm\!2.235$ & 0.832 & 38.42 & 17.26 & 4.56 \\
\midrule
SAGE & -- & -- & $6.245\!\pm\!1.960$ & 0.577 & 50.18 & \textbf{4.04} & 7.50 \\
MUSIC & -- & -- & $5.603\!\pm\!2.195$ & 0.619 & 44.85 & 4.54 & 3.13 \\
ESPRIT & -- & -- & $6.082\!\pm\!1.964$ & 0.581 & 57.66 & 5.57 & 4.90 \\
Joint OMP & -- & -- & $6.174\!\pm\!2.000$ & 0.581 & 52.08 & 4.06 & 6.88 \\
\bottomrule
\end{tabular}
\end{table*}

\begin{figure*}[!t]
\centering
\includegraphics[width=0.6\textwidth]{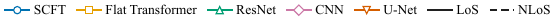}\\[-1.0mm]
\subfloat[]{\includegraphics[width=0.24\textwidth]{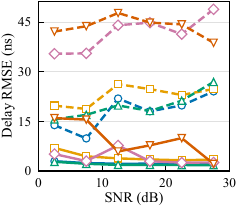}\label{fig:snr-delay}}
\hfill
\subfloat[]{\includegraphics[width=0.24\textwidth]{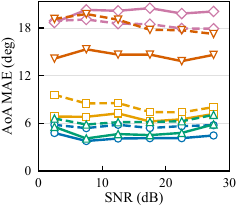}\label{fig:snr-aoa}}
\hfill
\subfloat[]{\includegraphics[width=0.24\textwidth]{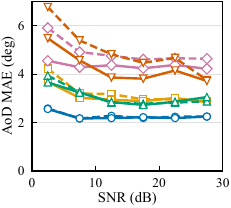}\label{fig:snr-aod}}
\hfill
\subfloat[]{\includegraphics[width=0.24\textwidth]{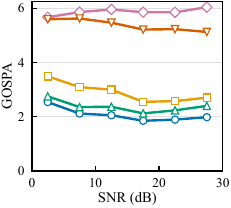}\label{fig:snr-gospa}}
\caption{MPC estimation performance versus SNR: (a) delay RMSE, (b) AoA MAE, (c) AoD MAE, and (d) GOSPA. Solid and dashed curves denote LoS and NLoS paths, respectively, in (a)--(c).}
\label{fig:snr_results}
\end{figure*}

\begin{table}[!t]
\centering
\caption{Controlled architectural ablations.}
\label{tab:ablations}
\scriptsize
\setlength{\tabcolsep}{3.0pt}
\begin{tabular}{lrrr}
\toprule
\textbf{Variant} & \textbf{GFLOPs} & \textbf{GOSPA} & \textbf{F1} \\
\midrule
\textbf{SCFT (spectral convolution)} & \textbf{3.621} & \textbf{2.001} & \textbf{0.942} \\
Full spectral attention & 3.777 & 2.351 & 0.938 \\
No cross-pair attention & 3.681 & 2.406 & 0.931 \\
No beam descriptors & 3.777 & 7.247 & 0.604 \\
\bottomrule
\end{tabular}
\end{table}

Table~\ref{tab:estimator_results} shows that SCFT achieves the lowest GOSPA and highest F1 among the evaluated methods. SAGE attains a lower AoA MAE on matched paths, despite substantially worse GOSPA and F1. The high aggregate delay RMSEs of SAGE and joint OMP mainly reflect NLoS errors (61.51 and 63.72 ns, respectively), whereas their LoS errors remain substantially lower (1.36 and 1.46 ns) on accepted matches. At 25--30 dB, their overall delay RMSEs remain 57.72 and 60.46 ns, respectively, compared with 20.78 ns for SCFT. For the learning-based methods, Fig.~\ref{fig:snr_results} shows that SCFT maintains the lowest GOSPA across the SNR range and consistently outperforms the other models in AoA and AoD estimation for both LoS and NLoS paths. However, its NLoS delay RMSE remains relatively high and comparable to ResNet, contrasting with the substantially smaller LoS errors. The GOSPA advantage thus reflects improved joint estimation rather than uniformly lower errors across individual parameters.

Table~\ref{tab:ablations} examines the contribution of each architectural component. Replacing local spectral convolution with full spectral attention increases GOSPA, with little change in F1. Starting from this full-attention variant, removing cross-pair attention further increases GOSPA, while removing the beam descriptors produces the largest degradation in both metrics. These results support combining local spectral processing with descriptor-conditioned interactions across beam pairs.

\subsection{Snapshot Positioning and Radio SLAM}

The downstream evaluation uses a separate 100-pose trajectory with 50 paired noisy channel realizations. The receiver clock bias is initialized uniformly over $[-5,5]$ ns and follows a Gaussian random walk with a standard deviation of $0.071$ ns per 2-s interval.

For independent snapshot positioning, SCFT and ResNet use the earliest detected direct path, while the oracle uses the earliest true direct path. Since SAGE does not predict path classes, its earliest detected path is treated as a direct-path candidate. The position is determined from the selected path's observed delay and global AoD, accounting for the known BS--UE height difference and assuming zero clock bias. Snapshots without an eligible path or valid range geometry are excluded. Snapshot RMSE is computed over the 4,980 cases, out of 5,000, for which all four methods yield valid positions.

For radio SLAM, a causal factor-graph backend jointly estimates 2-D position, heading, receiver clock bias, and vertical-facade virtual anchors using motion and MPC measurements. The delay measurement model accounts for clock bias. Direct-path candidates undergo innovation gating, while single-reflection paths are associated one-to-one with virtual anchors. Candidate anchors are confirmed after three distinct associated observations. Neither task uses path-gain or Doppler estimates. We also evaluate both tasks using oracle MPCs, providing ground-truth path classes and geometric parameters to the same positioning and SLAM backends. We report position RMSE, translational relative-pose error (RPE), virtual-anchor map GOSPA, and association F1.

\begin{table}[!t]
\centering
\caption{Snapshot positioning and radio SLAM results ($\mu\pm\sigma$). RMSE, RPE, and map GOSPA are in metres.}
\label{tab:downstream}
\scriptsize
\setlength{\tabcolsep}{2pt}
\renewcommand{\arraystretch}{1.05}
\begin{tabular}{@{}lrrrr@{}}
\toprule
\textbf{Metric} & \textbf{SCFT} & \textbf{ResNet} & \textbf{SAGE} & \textbf{Oracle} \\
\midrule
Snapshot RMSE & $\mathbf{2.55\!\pm\!0.16}$ & $3.13\!\pm\!0.14$ & $35.30\!\pm\!1.11$ & $0.78\!\pm\!0.41$ \\
SLAM RMSE & $\mathbf{1.78\!\pm\!0.37}$ & $5.09\!\pm\!7.93$ & $25.79\!\pm\!85.03$ & $0.83\!\pm\!0.32$ \\
RPE & $\mathbf{0.47\!\pm\!0.05}$ & $0.94\!\pm\!0.87$ & $21.86\!\pm\!119.15$ & $0.18\!\pm\!0.03$ \\
Map GOSPA & $4.94\!\pm\!1.12$ & $\mathbf{3.997\!\pm\!1.08}$ & $11.90\!\pm\!2.91$ & $1.26\!\pm\!1.33$ \\
Association F1 & $\mathbf{0.879\!\pm\!0.024}$ & $0.813\!\pm\!0.093$ & $0.081\!\pm\!0.038$ & $0.977\!\pm\!0.013$ \\
\bottomrule
\end{tabular}
\vspace{-0.1in}
\end{table}

\begin{figure*}[!t]
\centering
\includegraphics[width=0.8\textwidth]{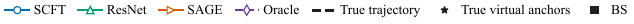}\\[-1.0mm]
\subfloat[]{\includegraphics[width=0.31\textwidth]{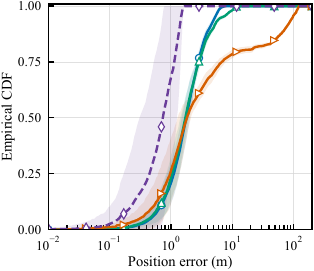}\label{fig:snapshot-cdf}}
\hfill
\subfloat[]{\includegraphics[width=0.31\textwidth]{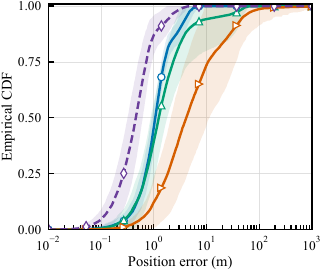}\label{fig:slam-cdf}}
\hfill
\subfloat[]{\includegraphics[width=0.31\textwidth]{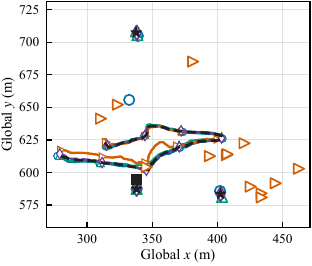}\label{fig:trajectory-map}}
\caption{Positioning and radio SLAM results: (a) snapshot positioning error, (b) radio SLAM position error, and (c) representative trajectories and virtual-anchor maps. Panels (a) and (b) show mean empirical CDFs with pointwise standard-deviation bands across 50 noisy realizations. Panel (c) shows a common realization selected based on proximity to the method-wise median SLAM RMSE and map GOSPA.}
\label{fig:downstream}
\end{figure*}

Table~\ref{tab:downstream} and Fig.~\ref{fig:downstream} show that SCFT achieves lower snapshot positioning errors than ResNet and SAGE. Although ResNet obtains a lower mean map GOSPA, SCFT achieves lower SLAM RMSE and RPE, higher association F1, and less variable trajectory errors. The snapshot-to-SLAM comparison reflects the combined effects of motion information, temporal filtering, clock and heading estimation, data association, and mapping constraints, rather than the contribution of NLoS mapping alone.

\section{Conclusion}
\label{sec:conclusion}

This paper presented SCFT for joint MPC estimation in mmWave MIMO-OFDM systems. By combining local spectral convolution, permutation-equivariant beam-pair attention, and a query-based decoder, SCFT enables multipath detection and joint estimation of geometric parameters, namely observed delay, AoA, and AoD, from noisy complex channel observations. Experiments on realistic ray-tracing data showed improved MPC set estimation over the evaluated baselines at a computational cost comparable to other learning-based methods.
Using SCFT's estimated MPCs for snapshot positioning and radio SLAM yielded lower positioning errors and more stable trajectories compared with the competing methods. Generalization beyond the considered environment, antenna configuration, and beam codebook remains untested and will be investigated in future work.

\balance
\bibliographystyle{IEEEtran}
\bibliography{main}

\end{document}